\documentstyle[11pt,epsf]{article}
\input psfig.sty
\def\beq{\begin{eqnarray}}
\def\eeq{\end{eqnarray}}
\def\bea{\begin{eqnarray*}}
\def\eea{\end{eqnarray*}}

\def\centeron#1#2{{\setbox0=\hbox{#1}\setbox1=\hbox{#2}\ifdim
\wd1>\wd0\kern.5\wd1\kern-.5\wd0\fi
\copy0\kern-.5\wd0\kern-.5\wd1\copy1\ifdim\wd0>\wd1
\kern.5\wd0\kern-.5\wd1\fi}}
\def\ltap{\;\centeron{\raise.35ex\hbox{$<$}}{\lower.65ex\hbox{$\sim$}}\;}
\def\gtap{\;\centeron{\raise.35ex\hbox{$>$}}{\lower.65ex\hbox{$\sim$}}\;}

\def\singleandthirdspaced{\baselineskip=\normalbaselineskip\multiply
    \baselineskip by 130\divide\baselineskip by 100}

\newcommand{\newc}{\newcommand}
\newc{\qbar}{{\overline q}}
\newc{\Kahler}{K\"ahler }
\newc{\deltaGS}{\delta_{\rm GS}}
\begin{document}
\begin{titlepage}
\begin{flushright}
{\large hep-th/yymmnnn \\ SCIPP-2009/09\\
}
\end{flushright}

\vskip 1.2cm

\begin{center}

{\LARGE\bf Axions in Gauge Mediation}

\vskip 1.4cm

{\large Linda M. Carpenter, Michael Dine, Guido Festuccia and Lorenzo Ubaldi}
%\\
\vskip 0.4cm
{\it Santa Cruz Institute for Particle Physics and
\\ Department of Physics, University of California,
     Santa Cruz CA 95064  } \\
%{\it $^b$Institute for Advanced Study, Princeton, New Jersey,  08540  }\\
%{\it $^c$Physics Department, Rutgers University, Piscataway, New Jersey}
\vskip 4pt

\vskip 1.5cm

\begin{abstract}
In supersymmetric theories, the presence of axions usually implies the existence of a non-compact,  (pseudo)moduli space.
In gauge mediated models, the axion would seem a particularly promising
 dark matter candidate.  The cosmology of the moduli then constrains the gravitino mass and the axion decay constant;
the former can't be much below $10$ MeV; the latter can't be much larger than $10^{13}$ GeV.
Axinos, when identifiable, are typically heavy and do not play an important role in cosmology.
\end{abstract}

\end{center}

\vskip 1.0 cm

\end{titlepage}
\setcounter{footnote}{0} \setcounter{page}{2}
\setcounter{section}{0} \setcounter{subsection}{0}
\setcounter{subsubsection}{0}

%%%%%%%%%%%%%%%%%%%%%%%%%%%%%%%%%%%%%%%%%%%
%%%%%%%%%%%%%%%%%%%%%%%%%%%%
\singleandthirdspaced

\section{Introduction}
The strong CP problem, as presently understood, requires either an axion, a massless $u$ quark, or spontaneous breaking of CP at some high energy scale.
Lattice gauge calculations appear to rule out a massless $u$ quark\cite{lattice}; at least in the widely explored landscape framework, the spontaneous CP solution seems
unpromising\cite{dinecp}.  So in many ways, the axion solution to the strong CP problem seems the most likely of the three to be realized in nature.  From
 astrophysics, there is a lower bound on $f_a$ of order $10^9$ GeV.  If cosmology
is conventional (in particular, if the universe was in thermal equilibrium up to some very high temperature, and not allowing for possible anthropic selection
effects), the axion decay constant is less than about $10^{13}$ GeV.  In unconventional cosmologies\cite{dfs,turner2,banksdinehv,banksdineaxion}, or allowing for anthropic selection\cite{lindeaxion,artw,freivogel}, it may be larger.

One can attempt to implement the Peccei-Quinn (PQ) solution of the strong CP problem in a variety of frameworks:  string theory, low energy effective field theory, and
models with or without supersymmetry.  Some of the issues involved in each of these possibilities are surveyed in \cite{dinecp}.  In this paper,
we focus on models of low energy supersymmetry, with gauge interactions as the messengers of supersymmetry breaking.  Unlike intermediate scale
(``supergravity'') breaking, there is no particular compelling dark matter candidate, so the axion seems worthy
of study.  In this context, the scale of
PQ breaking, $f_a$, is necessarily far larger than the scale of supersymmetry breaking.  So the axion lies in a supermultiplet, with a scalar field known
as the saxion, and a fermion, the axino.  As we will see, there is always a modulus in such models which determines the value of $f_a$, but this
modulus need not be the saxion.

In gauge mediated models, theoretical
considerations constrain the underlying scale of supersymmetry
breaking, $\sqrt{\vert F\vert}$, to lie roughly between about $10^5$ GeV and $10^9$ GeV.\footnote{One will sometimes hear it said that $\sqrt{\vert F \vert}$
as large as $10^{11}$ is allowed.  We are assuming here that the underlying microscopic theory does not even approximately conserve CP, or any
approximate flavor symmetries.}
The light modulus associated with PQ symmetry breaking is subject to significant cosmological constraints\cite{fischleretal,bkn}.  In this paper we will argue, first, that this particle
is unacceptable unless the axion multiplet couples directly to the messengers and/or the sector responsible for supersymmetry breaking.  If the axion
is to be the dark matter, even this is unacceptable
unless the scale of supersymmetry breaking lies to the {\it high} end of the allowed range, $10^{8}$ GeV or so.\footnote{Scales similar to those considered
here are an integral part of the $F$ Theory GUT program of \cite{vafa1,vafa2}.  The underlying pictures, however, are quite different, so this convergence of scales is remarkable.}

In the rest of this paper, we will explore models in which susy-breaking dynamics and the couplings to messengers determine $f_a$ and the mass of the
superpartners of the axion.
We will construct a variety of models which
illustrate the issues we have outlined above.  In one class, the saxion is a pseudomodulus; in another,
 the saxion is relatively heavy, with another field determining the value of $f_a$.  In these
models, we can explore the cosmological issues in a sharp fashion.  Another issue which will concern us is the origin
of the hierarchy between the PQ scale and the scale of susy breaking.  We will explore whether
this  ``PQ hierarchy'' might arise through small couplings or through tuning, or dynamically.

The rest of this paper is organized as follows.
In section \ref{raxion}, we note that gauge mediated models typically possess an approximate $R$ symmetry, which must be spontaneously
broken.  It is natural to ask whether the Goldstone boson of that symmetry, the $R$-axion, might be the QCD axion.
We explain why the answer is almost certainly no.

In section \ref{firstmodel}, we present models in which the Peccei-Quinn symmetry is broken at tree level, but there is a pseudomoduli space
on which $f_a$ varies.   In one example, the saxion is itself a pseudomodulus.  In a second, the ``$R$-saxion'' (the light modulus
in the Goldstino multiplet) is the pseudomodulus which determines $f_a$.  The axino and saxion are already massive
at tree level.  At one loop, the mass of
the pseudomodulus is estimated.
In section \ref{cosmology}, we discuss the cosmological constraints on gauge mediated models under the assumption
that the axion is the dark matter.  The most severe arises from coherent production of the lightest
pseudomodulus.
We demonstrate that in a generic inflationary model, the saxion starts a distance of order $f_a$ from its minimum.
As a result, it comes to dominate the energy density of the universe around the time of nucleosynthesis, with problematic
consequences for cosmology,  unless its mass is large enough.  This translates, in turn,
into a lower bound on the supersymmetry breaking scale as a function of $f_a$.  We argue that, as a result, in such a framework,
messenger masses are likely to be
of order $f_a$.  In a broad range of circumstances, the axino is short lived on cosmological time scales, and does not pose additional problems.
Moreover, from this consideration alone, $f_a$ cannot be too large; $f_a$'s significantly larger than $10^{13}$ GeV seem unacceptable.
If the axion is not dark matter, the cosmological
constraint arising from pseudomoduli is weaker or even non-existent.

In the models of section \ref{firstmodel},  the hierarchy between the supersymmetry-breaking scale and $f_a$ is simply put in by
hand, the result of a large ratio of masses in the underlying lagrangian.
One might hope to avoid this in either of two ways.  First, the parameters themselves might arise through ``retrofitting'', i.e. they might be dynamically
determined by other interactions.  Alternatively, and more economically,
given that the saxion is a pseudomodulus, with a potential varying only logarithmically
over large regions of the field space, it would
seem that such a potential could naturally have a minimum at a hierarchically large value of $f_a$.  We explore this possibility
in section \ref{hierarchy}, and find that this can be achieved only in a restricted set of circumstances.

Lurking in the background in all of this discussion is the question of whether unknown high energy effects spoil the Peccei-Quinn solution
of the strong CP problem.  As is well-known, this question can be organized in terms of higher dimension, CP-violating operators\cite{higherdimension}.  Even with
a relatively low scale of supersymmetry-breaking and Peccei-Quinn symmetry breaking, it is necessary to suppress operators up to rather high
dimensions.  Assuming that the Peccei-Quinn symmetry is an accidental consequence of a discrete $Z_N$ symmetry, for example, would seem
to require, in the models developed here, a huge value of $N$\cite{lazarides,dinediscretepq,heteroticorbifold,argentina}.   As discussed in \cite{dinecp}, {\it imposing} a constraint that
the axion constitute the dark matter, in these types of models, can {\it almost} account for the {\it quality} of the PQ symmetry.
If we do not require the axion be the dark matter, the required values of $N$ are more modest, but still rather large and now would simply
appear a peculiar accident.  We discuss this issue in section \ref{quality}.
In our concluding section, we summarize what we view as the principle lessons of this work, and
ask what features of the models considered here are likely to be robust.

\section{The $R$-axion as the QCD Axion}
\label{raxion}

A theorem of Nelson and Seiberg\cite{nelsonseiberg} asserts that generic theories which break supersymmetry possess an approximate $R$ symmetry.
In any realistic model, this must be spontaneously broken\footnote{It is necessary to break the $R$ symmetry to give mass to the gauginos.}.  The corresponding Goldstone boson is conventionally called the $R$-axion.
By way of nomenclature, we will refer to the pseudomodulus that accompanies the $R$-axion as the ``$R$-saxion''.

In any case, it is usually said that the $R$-axion gains mass when one couples the system to supergravity\cite{brp}.  This is because it is necessary,
in order to (nearly) cancel the cosmological constant, to insure that the expectation value of the superpotential takes a certain
value of order $m_{3/2} M_p^2$.  This is typically achieved by including a constant in the superpotential, which is then suitably tuned.
This constant breaks any would-be (discrete or continuous)
$R$ symmetry.  But one might hope, instead, that the $R$ symmetry is {\it spontaneously} broken by some dynamics\footnote{In a landscape,
there might be a large set of theories with a distribution of $\langle W \rangle$ accounting for the value of the cosmological constant.}

Upon careful consideration, however, this possibility does not appear promising.  The issue is a mismatch of scales.  In models of dynamical
breaking, one might expect that
\beq
\langle W \rangle = \Lambda^3~~~ \langle F \rangle = \Lambda^2
\eeq
In order to cancel the cosmological constant,
one needs, instead a hierarchy -- a huge hierarchy -- between the scale of $R$ breaking and the scale of supersymmetry breaking;
the scale of $R$ breaking, it would seem, should be of order the Planck scale, so
$f_a \sim 10^{18}$ GeV.  It is hard to see how this could be achieved consistent
with the constraints from cosmology and a high quality axion.

\section{Models With Hierarchical Breaking of the Peccei-Quinn Symmetry}
\label{firstmodel}

As we build models, we will first impose a global, continuous Peccei-Quinn symmetry.  Our viewpoint will be that this must eventually
be accounted for as an accidental consequence of an underlying discrete symmetry (or perhaps a continuous gauge symmetry).  One of us
has discussed this issue in the past\cite{dinediscretepq}, and we will comment further on this question in the concluding
section.

To outline the basic issues,
consider, first, a theory with fields $S_{\pm}$ carrying PQ charge $\pm 1$, and a neutral field, $\chi$.  Take for the superpotential:
\beq
W = \chi (S_+ S_- -\mu^2).
\label{unbrokensusy}
\eeq
Supersymmetry is unbroken in this theory.  The Peccei-Quinn symmetry is broken, and there is a moduli space, which can be understood in terms of the
``complexification of the symmetry group'':
\beq
S_{\pm} =\mu e^{\pm {\cal A}/\mu}.
\label{sphi}
\eeq
$f_a$ in this model is a function of ${\cal A}$
\beq
f_a = \sqrt{\vert S_+ \vert^2 + \vert S_- \vert^2}.
\eeq
${\cal A}$ is naturally referred to as the {\it axion supermultiplet}.
%{\bf ***One thing which bothers me is that as written $f_a$ is not holomorphic; probably a more
%natural definition of ${\cal A}$ absorbs $f_a$, i.e. $S_\pm = e^{\pm {\cal A}}$.***}

\subsection{Model In Which the Saxion is a Pseudomodulus}

We want to complicate this model so that supersymmetry is broken, at a scale much less than $\mu \sim f_a$.  We will consider, first, O'Raifeartaigh models, postponing
 (limited) discussion of dynamical breaking of supersymmetry and/or the Peccei-Quinn symmetry
 until later.  There are various strategies we might adopt.
In our first model,
we simply add modules to that of eqn. \ref{unbrokensusy},  which break supersymmetry, fix ${\cal A}$ ($f_a$), and act as messengers.    We will suppose that, in addition to the Peccei-Quinn symmetry, the model possesses an $R$ symmetry, as required by the theorem of Nelson and Seiberg\cite{nelsonseiberg} to break supersymmetry
in a generic fashion.  Examples of sectors which break supersymmetry and the $R$ symmetry are provided by the models of
\cite{shihr,dinemason,iss2}.    For definiteness, we take this sector to consist of fields $X$, with $R$ charge $2$, and $\phi_1,\phi_{-1}, \phi_3$, where the subscripts
denote the $R$ charges of the fields, and with superpotential
\beq \label{eq:mod1}
W_1 = \lambda_1 X(\phi_1 \phi_{ -1} - F) + m_1 \phi_1^2 + m_2 \phi_{-1} \phi_3.
\eeq
All of the fields in eqn. \ref{eq:mod1} are taken neutral under the $PQ$ symmetry.
As shown in \cite{shihr}, in this model, there is a range of parameters for which the $R$ symmetry is broken, and
\beq
\langle X \rangle = x + \theta^2 F.
\eeq
with $x \sim m_i$.   Note, in particular, that a hierarchy between $x$ and $F$ is possible.

We will take $S_\pm$ to have $R$ charge $0$.
In order to fix ${\cal A}$ in eqn. \ref{sphi}, rather than couple $S_\pm$ directly to $X$, we couple
these to another set of fields which
{\it do} couple to $X$.  The resulting model has a two dimensional (pseudo) moduli space, one direction associated with
breaking the $R$ symmetry, the other with breaking the $PQ$ symmetry.  The additional fields
$a_1,\bar a_1, a_2, \bar a_2, b_1, \bar b_1, b_2, \bar b_2$, {\it do not} transform under the gauge symmetries.  We take the
additional contribution to the superpotential to be
\beq
W_2 = \chi (S_+ S_- - \mu^2) + h S_+ a_1 \bar a_2 + y S_-  ~b_1 \bar b_2 + X( a_i \bar a_i + b_i \bar b_i)
\label{w2}
\eeq
We have not labeled the various independent couplings of $X$ to the $a_i$'s, and $b_i$'s to avoid cluttering formulas unnecessarily.  Also, to avoid cluttering the formulas,
we have not used the subscripts to label the $R$ or PQ charges, but this lagrangian is consistent with both symmetries.  For example, denoting $R$ and $PQ$
charges as $(R,PQ)$ we can take:
\beq
a_1 (1,-1), \bar a_1 (-1,1);\; a_2 (-1,0), \bar a_2 (1,0);\;b_1 (1,1), \bar b_1 (-1,-1);\; b_2 (-1,0), \bar b_2 (1,0).
\eeq
The superpotential of eqn. \ref{w2} is not the most general consistent with the symmetries.  For example,
we can add a term $\epsilon X S_+ S_-$; this is easily seen not to qualitatively alter the behavior of the model discussed below, at least for small $\epsilon$. Similarly for other possible terms.

In the model as it stands, the scales $f_a\sim \mu$ and $x \sim m_i$ are independent.  When we study
 cosmological issues in section \ref{cosmology}, however, we will see that cosmological considerations require that they be comparable.
We can write for $S_+$ and $S_-$:
\begin{equation}
S_+=\mu \sqrt{y\over h} e^{\phi\over \mu},\;\;S_-=\mu \sqrt{h\over y} e^{-{\phi \over \mu}}
\end{equation}
then in the limit of small $F$ and for $ \sqrt{h y} \mu x^{-1}$ not too large, the minimum of the one loop potential for the saxion (real part of $\phi$) is at $\phi=0$. The saxion acquires a mass
\begin{equation}
m_s^2\sim {1\over 16 \pi^2}{F^2\over \mu^2}
\end{equation}

The axino mass is parametrically lighter, by the square root of a loop factor.  We will see shortly that this is not generic; the axino can easily be -- and arguably
typically will be -- heavier than the lightest pseudomodulus.

We need one final module; it is important that the Peccei-Quinn symmetry be anomalous, and that supersymmetry breaking be transmitted
to the fields of the MSSM.  We accomplish this by coupling $X$ and $S$ to a set of fields filling out two $5$ and $\bar 5$'s of the
Standard Model:
\beq
W_3 = X (q_1 \bar q_1 + \ell_1 \bar \ell_1) + S_+ q_2 \bar q_2 + S_- \ell_2 \bar \ell_2.
\eeq
The fields $q_1, \bar q_1, \ell_1, \bar \ell_1$ act as messengers.  The fields $q_2$, etc., gain mass as a result of the
$S_\pm$ expectation values, giving rise to the coupling of the axion supermultiplet, ${\cal A}$, to the gauge fields of the Standard Model (MSSM).
Within this model, the masses of messengers and the scale $f_a$ are in principle independent.
One can contemplate more intricate messenger sectors, implementing general gauge mediation\cite{ggm, dcfm}.
But the most important point about this structure is that, given our argument in the next section that $f_a \sim x$,
messenger masses are likely to be of order $f_a$, in implementations
of the Peccei-Quinn symmetry within gauge mediation.

To summarize the spectrum, we have two pseudomoduli multiplets in this model. At the minimum of the Coleman-Weinberg potential:
\begin{enumerate}
\item  Superpartners of ordinary fields of the MSSM have masses given by standard model loop factors times $F/x$, as in conventional
gauge mediation.
\item  For $f_a \sim x$  the saxion mass is of order
\beq
m_s = \left ({1 \over 16 \pi^2} \right )^{1/2}{\vert F \vert \over f_a},
\eeq
 i.e. in terms of loop counting, larger than that of the MSSM
particles.
\item  The axino mass is the same loop order as those of the MSSM particles:
\beq
m_{\tilde a} \sim {1 \over 16 \pi^2} {\vert F\vert \over f_a}
\eeq
though it is further suppressed if $R$ symmetry breaking is small.  If $\vert x \vert \ll f_a$, the suppression is ${\cal O} (x/f_a)$.  In this
situation, in order to obtain suitable gaugino masses, there must be additional messengers which couple to $X$ but not to $S$.
\item  The ``r-axion" is massless, as is the Goldstino.  The ``r-saxion" gains mass at one loop similar
to that of the saxion.
\end{enumerate}

\subsection{Models in Which the ``R-Saxion" Determines $f_a$}

In the model of the previous section, there were two pseudomoduli, one responsible for determining $f_a$, one for determining $f_r$, the decay
constant of the ``r-saxion".  This arose because of our ``modular" structure:
by setting certain couplings
to zero, it was possible to decouple the sector which broke supersymmetry from the sector which broke the PQ symmetry.  In the absence of supersymmetry
breaking, the broken PQ symmetry implies the existence of a moduli space (associated with the ``complexification" of the symmetry group); the
requirement that supersymmetry is broken in the other, decoupled, sector, implies the existence of an $r$-axion, according to the theorem
of Nelson and Seiberg.
In this section, we construct models in which the symmetry-breakings cannot be decoupled.  The ``r-saxion" in these models determines
$f_a$; the saxion is not a pseudomodulus at all, and is   parametrically much more massive than the r-saxion.  The axino is massive already at tree level,
so much more massive than the lightest pseudomodulus.

The model, again, contains fields $S_\pm, \chi$, with quantum numbers as before.  In addition, there are fields $X,Y$ with $R$-charge
$2$ and PQ charge $\pm 2$, respectively.  There are additional fields, $\phi_1,\phi_{-1}$ with the following  $(R,PQ)$ charges:
\beq
\phi_1(1,0),~ \phi_{-1}(-1,-2);~\tilde \phi_1(1,0),~\tilde \phi_{-1}(-1,2).
\eeq
For the superpotential of the model
we take:
\beq
W = \chi (S_+ S_- -\mu^2) + X (\phi_1 \phi_{-1} -\lambda_1 S_-^2 ) + Y (\tilde \phi_1 \tilde \phi_{-1}- \lambda_2 S_+^2) + m \phi_1^2 + \tilde m \tilde \phi_1^2.
\eeq
Again, we have set some couplings to one to avoid cluttering formulas; our discussion is readily modified if these are allowed to vary.  This
superpotential {\it  is} the most general (renormalizable) one consistent with the global symmetries and a discrete $Z_2$ symmetry under which the tilde fields change sign.  Consider the limit of
very small $\lambda_1=\lambda_2 = \lambda$, $\mu \sim m$.  In this limit (assuming that $X,Y \sim m, \tilde m$), one linear combination of $S_\pm$ combines with $\chi$
to form a massive field.  The other linear combination is fixed, but lighter.  Classically, there is a one parameter moduli space, satisfying $X-Y = 0$.
The modulus is the field ${1 \over \sqrt{2}}(X+Y)$; this is the R-saxion.  The R-axion is the imaginary part of this field, and the Goldstino is the corresponding
fermion.  For large $X+Y$,  the field $X-Y$ is the saxion supermultiplet.  Because of the constraint, it is not a pseudomodulus, and indeed,
the real scalar and the axino are massive already at tree level, with mass of order $\lambda \mu$.
The one loop Coleman-Weinberg calculation yields a stationary point for the modulus, $X+Y$.
Its mass is of order
\beq
m_{\tilde r}^2 = {1 \over 16 \pi^2} {F^2 \over x^2}
\eeq
where $x \sim m, \tilde m$, and $F \sim \lambda \mu^2$.
In this region of the parameter space, the hierarchy between $F$ and $f_a$ is determined by the small couplings $\lambda$.

It is interesting to write the effective field theory for small $\lambda$, integrating out the massive fields.
Writing $S_+=\mu+\delta S_+,\;\;S_-=\mu+\delta S_-$ the massive fields (mass $\sim \mu$) are
\beq
\chi,\; H= {1 \over \sqrt{2}} (\delta S_+ + \delta S_-).
\eeq
This leaves the light fields ($m=0$ or $m \sim \lambda \mu$):
\beq
Z= {X + Y \over \sqrt{2}}~~~~ A = {\delta S_+ - \delta S_- \over \sqrt{2}}~~ B = {X - Y \over \sqrt{2}}.
\eeq
$Z$ is the Goldstino supermultiplet.  Its fermion is the goldstino; the imaginary part of the scalar is the $R$-axion, and the real part the $R$-saxion (the lightest pseudomodulus).
To obtain the effective lagrangian for the light fields, one needs to solve the heavy field equations of
motion.
Solving the equation $\partial W/\partial H=0$ yields $\chi= \sqrt{2} \lambda Z+O(\mu^{-1})$,
which substituted back in $W$ gives
\beq
W = -\lambda(\sqrt{2} \mu^2 Z + \sqrt{2}  Z A^2 - 2 \mu A B)+O(\mu^{-1}).
\eeq
This is an O'Raifeartaigh model.  It possesses an $R$ symmetry, but the PQ symmetry is explicitly broken in the effective theory.
The masslessness of the axion, at the level of the low energy lagrangian, appears as a consequence of tuning of the parameters. For small $Z$, the axion is principally
$A$, while for large $Z$, as expected, it is principally $B$.  There is no light axino.

For small $Z$, the saxion is a massive field, with mass of order $\lambda^2 \mu^2$.  For large $Z$, the field $A$ is heavy, while $B$ is light, with mass of order
$\lambda^2 \mu^4/Z^2$.   The real part of $Z$, the $R$-saxion, is still lighter by a loop factor.  $B$ is naturally described as the saxion; it is not a pseudomodulus.
For intermediate values of $Z$, the saxion and axion are different linear combinations of $B$ and $A$, and so are not precisely aligned; there is not a sharp notion of what
one means the by the saxion in these cases.

To generate the coupling of the axion to $F \tilde F$, it is necessary to couple the fields $X,Y$ to messengers.
A coupling of the form $
X q \bar q$,
for example, generates a coupling of both the axion multiplet and the r-axion multiplet to $W_\alpha^2$.  This is relevant
not only for the axion, but for the light modulus (moduli).  We will discuss the cosmological issues raised by these
models in the next section.
As in conventional gauge mediation, there are one loop diagrams contributing mass for gauginos and two loop
diagrams contributing mass to squarks and sleptons.

So far, we have not contemplated coupling $X$ to fields other than messengers, so one might think that the quarks and leptons should be
neutral under the PQ symmetry.  However, once one attempts to solve the $\mu$ problem, some coupling of messenger fields to fields
which couple to ordinary fields is required.  So generically, one expects that the quarks, leptons and Higgs will carry PQ charges.
Knowledge of these charges is required to fully determine the couplings
of the axion to ordinary fields; this bears on the question of axion  detection.

\section{Saxion/Axino/Axion Cosmology}
\label{cosmology}

A priori, the mass scale of the messengers responsible for transmitting supersymmetry breaking, and $f_a$, are independent.
We have asserted, however, that in practice the relation between these scales is subject to cosmological constraints.  In this section,
we treat the scale of the messengers (and thus the scale of supersymmetry breaking) and $f_a$ as independent, and make a preliminary examination of the consequences.
We will see that over much of the interesting parameter range, unless the messenger scale is within two orders of magnitude of $f_a$, the pseudomodulus
is very long lived and dominates the energy density shortly after nucleosynthesis.  

In models of the type we are considering here, the most serious cosmological issues are posed by the lightest modulus, which
we will denote by $P$.  Recall that the mass of $P$ is of order
\beq
m_P = \sqrt{\rm loop~factor} {F \over f_a} \sim 10^{-3} {\rm GeV}\left ( {F \over 10^{10} {\rm GeV}}\right )\left ( {10^{12} {\rm GeV} \over f_a} \right )
\eeq
($m_P$ is the mass of $P$, as opposed to the Planck mass, which we denote
$M_p$).

During inflation,
there is no reason for the lightest pseudomodulus to sit at the minimum of its potential; indeed, we expect it to sit a distance of order $f_a$
from its flat-space, zero temperature minimum.  For example, in the first of the models we considered,
during inflation, one expects that the
potential contains these additional terms (and many others):
\beq
\delta V = (a \vert S_+ \vert^2 + b \vert S_- \vert^2)H^2
\eeq
where $H$ is the Hubble constant during inflation; if $H < f_a$ the
the potential has a minimum at
\beq
{\cal A}= P \approx {\mu \over 4} \ln(b/a).
\eeq
Similar remarks hold for the second model (in which the saxion is not the pseudomodulus).
 In the class of models where the saxion is a pseudomodulus, $P$ decays principally to an axino and a gravitino; in the class where the lightest modulus
is the partner of the Goldstino, $P$ decays to a gravitino pair.  In either case, the decay width is of order
\beq
\Gamma_P  = {1 \over 16 \pi} {m_P^3 \over f_a^2} \sim 10^{-34}\left ( { 10^{12} {\rm GeV}\over f_a}\right  )^2\left ( { m_P \over 10^{-3} ~{\rm GeV}}\right  )^3 {\rm GeV}.
\eeq
After inflation, $P$ will be frozen at this point until $H \sim m_P$, at which time
$P$ starts to oscillate about its minimum.  $P$'s energy density at that time is of order $m_P^2 f_a^2$.  It constitutes a fraction of the energy density of order $f_a^2/M_p^2$.  It behaves like matter, so its fraction of the energy density grows with the scale
factor.  So, e.g., if $f_a = 10^{12}$ GeV, it represents a part in $10^{-12}$ of the energy density initially.   If its mass is, say, $10^{-3}$ GeV,  it starts to oscillate when $T \sim 10^{7.5}$ GeV, and dominates the energy density for $T \sim 30$ KeV.  Its lifetime is of order $10^{10}$ seconds, so there is a long period of matter domination before the normal time of matter-radiation equality.  This is followed by a {\it long} period in which the universe is dominated by relativistic gravitinos.  If the mass is closer to $1$ GeV, on the other hand, the decay occurs before nucleosynthesis, just when the modulus is coming to dominate the energy density. A significant fraction of the decay products will include hadrons, so nucleosynthesis will be problematic.

So quite generally, we seem to want the saxion mass (for $f_a \approx 10^{12}$ GeV) to be significantly greater than $1$ GeV.  As we have remarked above, if all of the messengers
have masses of order $f_a$, the saxion will readily satisfy this bound.  The situation is potentially problematic, however,
if some messengers are light compared to $f_a$.  For  $f_a = 10^{12}$ GeV and the loop
factor $10^{-3}$, for example, one requires $\sqrt{F} > 10^{7.5}$ GeV.  This is not much below the scale required if the messenger mass is of order $f_a$ ($\sqrt{F} = 10^{8.5}$
GeV),
so it seems most reasonable to suppose that the messengers have scale $f_a$.

For $f_a \sim 10^{9}$ GeV, there is no significant cosmological constraint arising from the pseudomodulus.

For many of the models we consider here, the axino is quite massive, and not problematic (or even interesting) cosmologically.  This seems
likely to be generic if $f_a$ is comparable to the messenger scale.  Conceivably there are situations where the axino mass
might be suppressed, and then the axino may
significantly constrain, e.g. the reheating temperature after inflation\cite{covi}.
Under such circumstances, it could constitute a significant fraction of the dark matter.

The cosmology of the pseudomoduli requires a more thorough analysis than we have presented here.  There may well be windows of low messenger scale which are cosmologically
allowed.  But it would seem from this discussion that the most elegant possibility is that the PQ scale is comparable to the messenger mass.  This does preclude, however,
very low scale gauge mediation.

\section{Obtaining a Large Hierarchy Dynamically} \label{hierarchy}

In the model of the previous sections, a large ratio of scales in the underlying lagrangian lead to a large hierarchy between the $f_a$ and $\sqrt{F}$. This could arise dynamically
in a model with ``retrofitted'' parameters\cite{retrofitted}.  One of us will explore this question
 elsewhere\cite{carpentertoappear}, but one might hope that it could arise in a somewhat different way.
Given that the saxion is necessarily a pseudomodulus, and that its potential varies only logarithmically for large values of the modulus, PQ breaking would seem highly
susceptible to such large hierarchies.  In this section, we explore strategies to models of this type.

\subsection{Models without New Gauge Fields}
To build simple models, one might introduce a field, $S$, carrying PQ charge and coupled to messengers.  Classically, the potential in the $S$
direction vanishes. The coupling to messengers leads to a potential for $S$ at one loop.  (Similar models, but in which the $S$ field is not a pseudomodulus, have been
considered in proposals to understand the $\mu$ term \cite{giudice1,giudice2}).  Specifically, we write \beq X (\lambda_1 M_1 \bar M_1 + \lambda_2 M_2 \bar M_2) + y S M_1
\bar M_2 \eeq where $X$, as before, is a field with $\langle X \rangle = x + \theta^2 F$, and the $M$'s are messengers, taken to transform, say, as $5 + \bar 5$ of $SU(5)$. This
model admits a Peccei-Quinn symmetry with $SU(3)$ and $SU(2)$ anomalies (under which $X$ transforms as well), so the dynamics responsible for supersymmetry breaking
must respect this symmetry.  One can take as transformation laws: \beq S \rightarrow e^{i \alpha} S; M_1 \rightarrow e^{-i \alpha} M_1; ~X \rightarrow e^{i \alpha} X, \eeq all
other fields being neutral.

Consider, now, the potential for $S$. At small $S$, as discussed in \cite{giudice1,giudice2}, the quadratic terms vanish.  But we are interested in the behavior at large $S$.  This
behavior is logarithmic.  If the coefficient of the log were negative, and if there were higher dimension terms which respected the symmetry, then one might get a hierarchically
large breaking of the PQ symmetry.  The coefficient of the logarithmic term is easily found to be:
\beq
 V(S) ={y^2 \over 16 \pi^2} \log(S^\dagger S) F^\dagger F.
\eeq
So we do not find the desired behavior.  This is readily seen to be quite general, in a theory with chiral fields only. The relevant contribution to the potential at second order in $F^2$ is:
\beq
V = {1 \over 16 \pi^2} \int d^4 \theta
{\rm Tr} \left (M^\dagger M \log (M^\dagger M) \right )
\eeq
 the computation simplifies, as we are only looking for the logarithmic behavior for large values of the fields.
One can always define a single field, $X$, to have an $F$, component, with other fields, $\phi_i$, having zero F components. Then it is straightforward to show that the term in
${\rm Tr}~M^\dagger M$ proportional to $X^\dagger X$ is always positive. Allowing for large $x$, corresponding to a large breaking of the $R$ symmetry, does not alter the
situation.  Such models do not produce the desired hierarchy. The same remains true in the models discussed by Shih\cite{shihr}, in which there are fields with $R \ne 0,2$,
which yield $R$ symmetry breaking. The potential still grows logarithmically for large values of the fields.

As an alternative, one can consider models (again like those of \cite{shihr} which exhibit runaway behavior in certain regions of the field space.  One might then hope to stabilize
the field through higher dimension operators.  The difficulty with this approach is that typically the higher dimension terms give rise to a supersymmetric minimum.  This is not
surprising.  In the runaway directions, supersymmetry becomes better and better as the fields become larger. As follows from the analysis of\cite{dcfm}, the effective theory with
the higher dimension operator typically has a superpotential of the type: \beq {1 \over \phi^n} + {\phi^{n+3} \over M_p^n}. \eeq This has a supersymmetric minimum.
Stabilization can be achieved in a {\it local} minimum if, say, the leading allowed operator has the form \beq {1 \over \phi^n} + {X\phi^{n+2} \over M_p^n} \eeq for some other
field $X$.

\subsection{Models with Additional Gauge Interactions}

With additional gauge interactions (beyond those of the Standard Model), one might hope to find runaway behavior in the Coleman Weinberg calculation.  It is well-known that
the gauge corrections to the potential tend to be negative for large values of charged fields.  If some of the charged fields also carry Peccei-Quinn charge, one might obtain the
desired structure.  The potential, at large fields, might be stabilized by higher order corrections (understood through the renormalization group, as in
\cite{inverted,bankskaplunovsky}) or through higher dimension operators in the superpotential.  We have found that it is possible to realize this possibility, but that the
resulting models are rather complicated.

Take a simple example, the model of ref. \cite{dinemason,iss2}, with fields $\phi_\pm, Z_\pm$, etc.  Add fields $S_\pm^{(i)}$, $i=1,2$, which are assumed to transform, as well,
under a PQ symmetry: \beq S_{\pm}^{(i)} \rightarrow e^{\pm i\alpha} S_\pm^{(i)} \eeq For the superpotential, we take \beq W = \lambda Z_0 (\phi_+ \phi_- - \mu^2) + M_1 Z_+
\phi_- + M_2 Z_- \phi_+ \eeq The $S$ fields do not appear in the superpotential at the renormalizable level.  For sufficiently large $\mu$, $\phi_\pm$ obtain expectation values.
If $\lambda$ is not too small, the fields $Z_0,Z_\pm$ obtain vanishing expectation values.  The fields $S_\pm^{(i)}$ can obtain large expectation values, constrained only by the
$U(1)$ D term.  Adding higher dimension operators of the type \beq \delta W ={\gamma_i\over M^{(n-3)}}(S_+^{(i)} S_-^{(i)})^n \eeq respects both the gauge and PQ symmetries, and
can stabilize the potential. Both symmetries are spontaneously
 broken by large amounts (proportional to a fractional power of $M$).  There are a number of fields, in addition to those in the axion multiplet,
with masses of order $F/f_a$, times loop factors.  It is necessary, of course, to couple the $S^{(i)}$ fields to messengers, and to impose, say, discrete symmetries which account
for this structure.  The models are quite baroque, but at least provide an existence proof that this sort of model building is possible.

An alternative, closer in spirit to Witten's inverted hierarchy\cite{inverted}, invokes a new, non-abelian ($SU(2)$) gauge group, Consider the following superpotential
\begin{equation} W=\lambda X( Tr[\phi_+ \phi_-]-\mu^2)-m Tr[Z \phi_-]-\alpha Y Tr[\phi_+^2] \end{equation} The fields $\phi_{\pm},\;\;Z$ are in the adjoint of $SU(2)$ while
$X$ and $Y$ are singlets.  The PQ charges of $\phi_{\pm}$ are $\pm 1$, while $Z$ and $Y$ have PQ charge $1$ and $-2$ respectively. $\mu$ and $m$ are mass parameters.
$X,\;\;Y,\;\;Z$ have $R-$charge 2 while $\phi_{\pm}$ have 0 $R-$charge.

At the minimum of the potential all the adjoints fields commute and their vev's can be simultaneously diagonalized. $SU(2)$ is broken to $U(1)$ and the pseudomoduli space is
three dimensional and includes the axion and the R-axion. Exploiting the $R$ and $PQ$ symmetries to set $X,Y\in R$. We have: \begin{eqnarray} Z&=& Y \sigma_3,\;\;Y=X {s
\lambda \over 2 \alpha}\\ \phi_-&=& s \phi_+,\;\;\phi_+={ s m\over 2 \alpha }\sigma_3 \end{eqnarray} where $s$ is determined in function of the parameters of the model as
follows: \begin{equation} s (2\alpha^2+ s^2 \lambda^2)={2 \alpha^2 \lambda^2 \mu^2 \over m^2 } \end{equation} The potential at the minimum is \begin{equation}
V=|F|^2={s^2 (4\alpha^2+3 s^2 \lambda^2)\over 4 \alpha^2 \lambda^2}m^4 \end{equation}

For large $X$, the tree level spectrum is comprised of: \begin{enumerate} \item Fields which are mostly combinations of $\phi_+,\;\phi_-$ with masses of order $\lambda X$
\item The rest of the fields charged under the unbroken $U(1)$ with masses of order $g Y$ \item
Massless particles:   three massless scalars: the axion, R-axion and R-saxion; the
goldstino; one massless gauge boson for the unbroken $U(1)$ with its gaugino. \item All the remaining fields neutral under the unbroken $U(1)$ get masses of order
${\mu^4\over X^2}$ \end{enumerate}
The R-saxion and the $U(1)$ gaugino receive masses by quantum
effects which we estimate in the following.

For $X\gg \mu,\;m$ the R-saxion potential dependence on $X$ is obtained by replacing the parameters in the tree level potential with their values at the scale $X$. The RG
equations are such that for ${\alpha^2 }>{4\over 3}g^2$ the effective potential is increasing. There is an $IR$ fixed point at $\alpha^2={3\over 10} \lambda^2={12\over 31}g^2$
while in the UV  $\alpha$ and $\lambda$ are either asymptotically free or both diverge. In the latter case by choosing the initial conditions with $\alpha^2<{4\over 3}g^2$ and
decreasing potential a minimum will be generated for larger values of $X$.

It is not hard to achieve the required hierarchy between $X$ and $\sqrt{|F|}$ with all the couplings
remaining well in the perturbative regime for the range of $X$ of interest. One example of the resulting potential and RG running is shown below for $g=1$.   The R-saxion typically
acquires a mass of order ${1\over (4\pi)}{F\over X}$ as does the gaugino for the unbroken $U(1)$.

\begin{figure}[h!]
\centering
 \begin{tabular}{ccc}
\psfig{figure=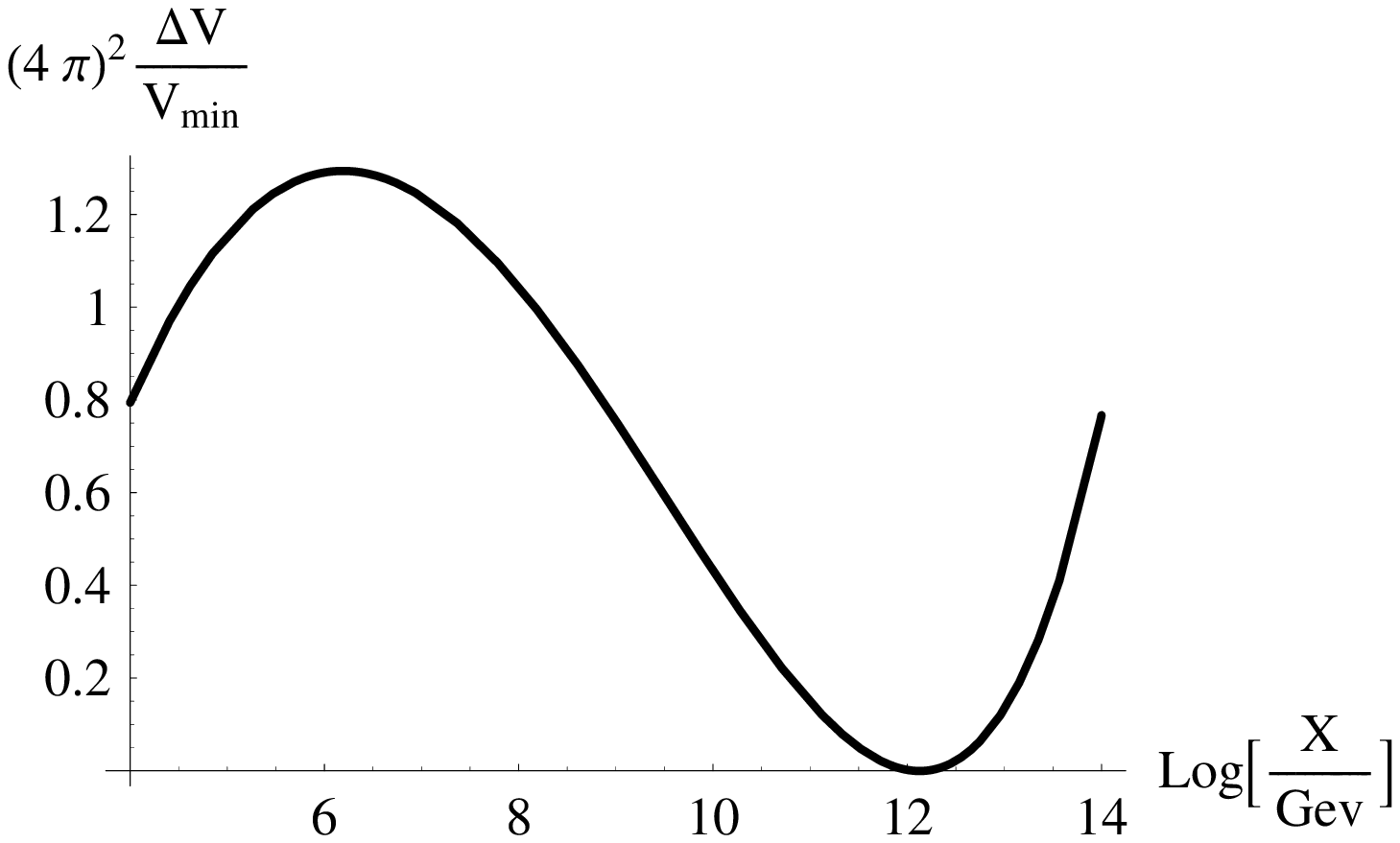,width=2.2in}
\psfig{figure=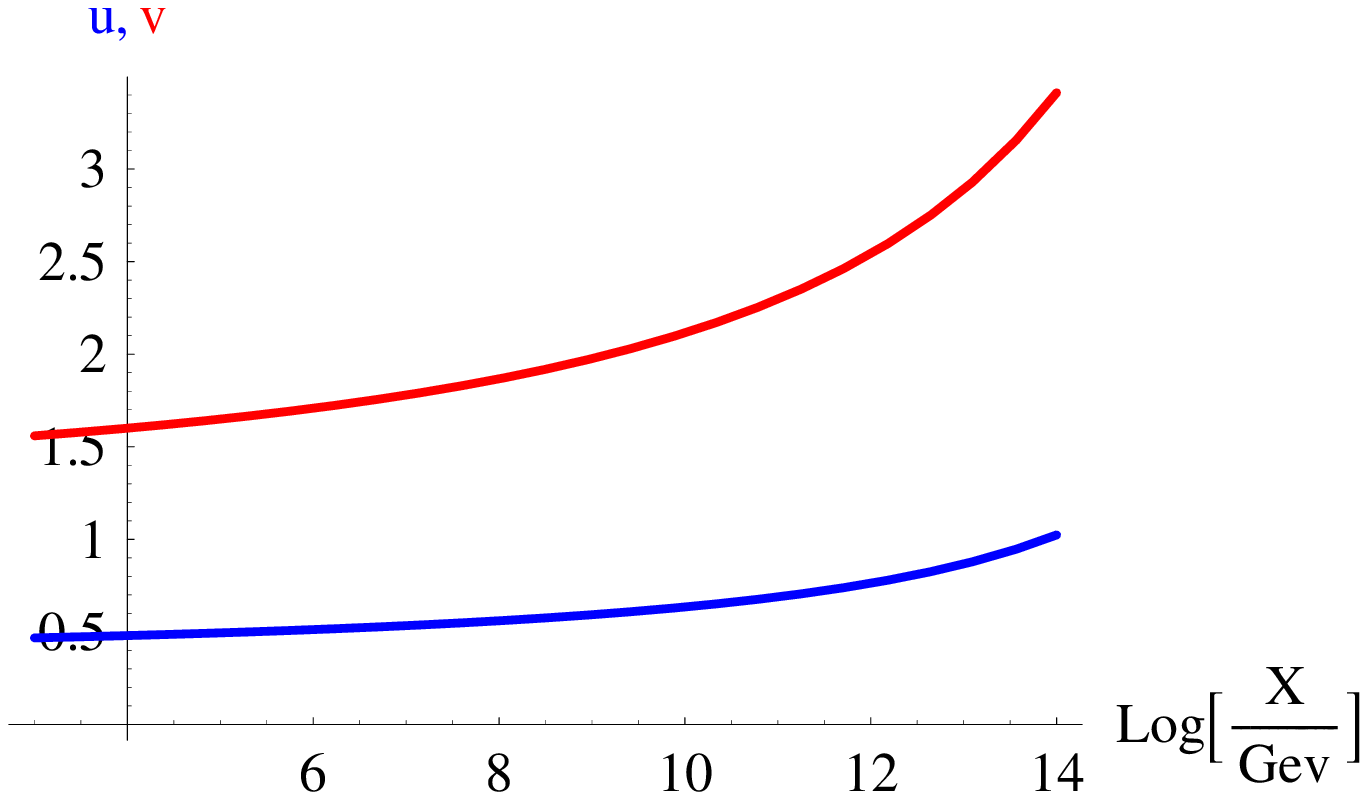,width=2.2in}
\psfig{figure=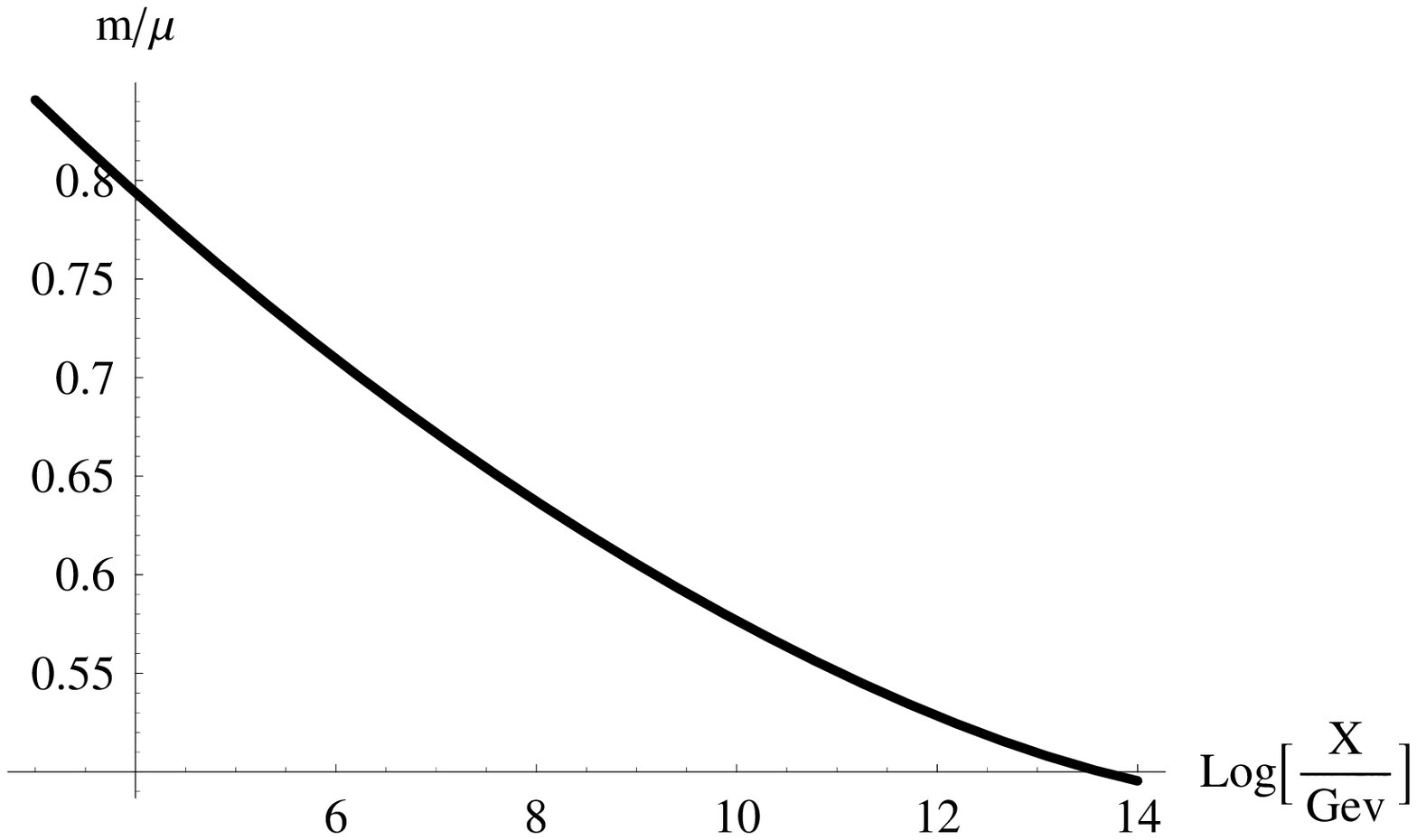,width=2.2in}
\end{tabular}
\caption{Plot for the potential in function of $\log(X)$. $\Delta V= V(X)-V(X_{min})$. Also plotted are $u={\alpha^2\over g^2} ,v={\lambda^2\over g^2}$ and ${m\over \mu}$ as a function of $\log(X)$}
\end{figure}

We can now couple the singlet $Y$ to messengers in $\bf{5},\;\bf{\bar{5}}$ of $SU(5)$. Provided that the coupling to the messengers is not too large, the potential will still develop a minimum.  There is a small kinetic mixing $\epsilon$ arising from 3 loop effects between the hidden unbroken $U(1)$ and the $EM\;\;U(1)$.
 As a result  all of the fields charged under the former acquire $EM$ charges of order $\epsilon$.
 As all fields charged under the unbroken $U(1)$ have masses of the order of the PQ breaking scale $f_a$ in order to avoid
cosmological constraints\cite{Holdom:1986eq,Davidson:1991si} we need $f_a> \epsilon^2 10^{15}$Gev.

This model indicates that it is possible, in principle, to account for the large value of $f_a$ relative to $\sqrt{\vert F \vert}$ through an inverted hierarchy.  But the model is
complex, and introduces new, possibly undesirable, light degrees of freedom (such as the massless gauge bosons of the extra $U(1)$).  In the end, models such as those we have
discussed which account for the PQ hierarchy through small parameters are perhaps more plausible.

\section{Axion Quality}
\label{quality}

The most troubling feature of the axion solution to the strong CP problem is the requirement that the PQ symmetry
be of extraordinarily high {\it quality}.  In \cite{dinecp}, the quality, $Q$, was defined by writing the axion potential as
\beq
V_a =Q f_a^4 \cos({a \over f_a} - \theta_0).
\eeq
In order that the axion solve the strong CP problem, one requires $Q< 10^{-62}$, if $f_a = 10^{12}$ GeV.

Adopting the language of the ``landscape", states (or theories) with such a high quality PQ symmetry
are not likely generic; one can ask what might select for them.  Since the laws of nuclear physics are hardly sensitive
to $\theta$, the only plausible answer we can see is  that there might be classes of theories (states) for which
axions are a generic form of dark matter, and that something close to the presently observed dark matter density
is a requirement for a hospitable universe\cite{lindeaxion,artw,freivogel}.
In string theory (and/or in higher dimension theories) we understand how a
non-linearly realized PQ symmetry can arise in the four dimensional theory as an accidental consequence
of features of the microscopic theory.  As explained in reference \cite{dinecp}, selecting for dark matter in such settings
may well account for the requisite $Q$.

In gauge mediation the situation is different.
As we have stressed here, and as discussed also in \cite{dinecp},  if gauge mediation is the origin of supersymmetry
breaking, and an axion is responsible for the absence of strong CP violation, then a sensible
cosmology requires that any ``stringy moduli" be fixed at very high energies (supersymmetrically), and
the Peccei-Quinn symmetry must be broken within
the low energy field theory, as in the models discussed in this paper.
Within field theory models, the sensitivity to unknown high energy
effects is particularly troubling\cite{higherdimension}:   in some range of energy, the theory must exhibit
a {\it linearly} realized global symmetry of extremely
good quality.

%Even with comparatively low scale supersymmetry breaking, the problems are severe.  The problem is that
%we require that any Peccei-Quinn violating effect contribute to the axion potential an amount no more than $10^{-10}$ or so of the QCD contribution,
%or about $10^{-14}$ GeV$^4$.
To appreciate the severity of the problem, note that
PQ violating operators can arise in the superpotential
\beq
\delta W = \phi^3 \left ({\phi \over M_p} \right )^{n-3}
\eeq
yielding contributions to the potential of order
\beq
f_a^2 \vert F \vert \left ({f_a \over M_p} \right )^{n-3} < 10^{-14} ~{\rm GeV}^4.
\eeq
In the present paper, we have argued that $F \sim 10^{17}$ GeV$^2$ is a natural scale; correspondingly we require
$n>13$!  Ignoring the cosmological issues we have raised, even with $f_a = 10^9$ GeV and the smallest $F$'s we can contemplate in gauge mediation, we require
$n>9$.   If the Peccei -Quinn symmetry results from, say, an underlying discrete $Z_N$ symmetry, we need $N>9$ or $N>13$ to have a viable
axion solution to the strong CP problem.

In a landscape framework, one might expect that such large discrete symmetries are extremely
rare\cite{dinesun}.     On the other hand, in gauge mediation, the axion is a particularly plausible dark matter candidate.  Again, as described
in \cite{dinecp}, selecting for such states gives a very flat potential, but not quite flat enough to solve the strong CP problem, except
for large $f_a$.  For $f_a = 10^{12}$ GeV, for example, selecting for axion dark matter
requires $n > 11$.  In other words, the dark matter requirement comes close, but whether
this is close enough is not clear.
(With supersymmetry broken at an intermediate
scale as in ``gravity mediation", dark matter does not seem a persuasive criterion, as there are other dark matter candidates which seem
less expensive.\footnote{Some discussion of axions in gravity
mediation, from a different point of view,
appears in \cite{baer}.} )

One might argue that a non-generic superpotential, perhaps plausible due to the non-renormalization theorems and to
experience with string theory\cite{evaed}, could account for the absence of such superpotential couplings.
However, significant PQ violating terms are likely to appear in the Kahler Potential,
\beq
K= \int d^4 \theta \phi \phi^{\dagger} \left ({\phi \over M_p} \right )^{n}
\eeq
and hence yield a constraint
\beq
 \vert F \vert \vert F^{\dagger} \vert \left ({f_a \over M_p} \right )^{n} < 10^{-14} ~{\rm GeV}^4.
\eeq
In our standard case where $F \sim 10^{17}$ GeV$^2$ and $f_a = 10^{12}$ GeV, we must have $n>8$.  Looking at the Kahler potential we see that this corresponds to a $Z_N$ of 8, a modest improvement over the superpotential constraint.

\section{Conclusions}

One of the great successes of (critical) string theories is that they yield Peccei-Quinn symmetries which hold to a high degree
of accuracy\cite{wittenpq,dinebook}; these PQ symmetries can often be thought of as an accidental consequence of higher dimensional
gauge symmetries. They are good symmetries to all orders of perturbation theory, but fail non-perturbatively.
Provided that the high energy non-perturbative effects are sufficiently small, these theories seem a suitable setting to implement
the axion solution of the strong CP problem.  As discussed in \cite{dinecp}, one faces at least two issues with such models.  First,
$f_a$ is likely to be large; one needs to reconcile this with cosmological constraints, perhaps along the lines of \cite{dfs,turner2,banksdinehv,banksdineaxion}
or \cite{lindeaxion,artw,freivogel}.  Second, one has to ask why, once moduli are fixed, there is an axion sufficiently light (a PQ symmetry of sufficient {\it quality})
to solve the strong CP problem.   As discussed in reference \cite{dinecp},one possibility  -- perhaps the only one --
is that there is a selection for axion dark matter.  For $f_a > 10^{14}$ GeV, account for the quality of the QCD axion.
If low energy supersymmetry plays no role in nature, or with intermediate scale breaking supersymmetry breaking,\footnote{As we have remarked,
in the intermediate scale case, it may be hard to understand why the axion is a more generic form of dark matter
than the conventional neutralino.} this might provide
an adequate understanding of the strong CP problem.  In gauge mediation,
we have seen that the requirement of axionic dark matter selects for a very good PQ symmetry, but perhaps not quite good enough (unless, again,
$f_a$ is uncomfortably large).

We can ask what features of the model of eqn. \ref{firstmodel} might be expected to be generic.  Among these:
\begin{enumerate}
\item  $f_a$ not much larger than $10^{13}$ GeV.  This seems forced by the challenges of saxion cosmology; otherwise, the decay of the saxion is very late, for any
plausible scale of supersymmetry breaking.
\item  Messenger masses of order the Peccei-Quinn scale:  again,  this seems forced by the challenges of saxion cosmology.  If another solution
is found to this problem, these conditions might be relaxed.
\item  Axinos with mass typically as large or larger than that of the lightest
pseudoscalar:  this seems generic, as a consequence of the first point,
which forces the $R$ symmetry breaking scale to be comparable to $f_a$.
\end{enumerate}

In the end, the axion solution to the strong CP problem, within the framework of gauge mediation, seems highly constrained.  The Peccei-Quinn
and messenger scales are likely to be similar.  If supersymmetry is discovered at the LHC, and if evidence accumulates for a gauge-mediated
supersymmetric spectrum, this could well point to a detectable axion as the dark matter, with the NLSP decaying well outside
of the detector.

\noindent
{\bf Acknowledgements:}
We thank David Shih for a number of insightful observations and an important question about gaugino masses.  Conversations
with Shamit Kachru, Nathan Seiberg, Steve Shenker and Leonard Susskind were of great value to us.  This work
supported in part by the U.S. Department of Energy.

\end{document}